\documentclass[journal]{IEEEtran}
\ifCLASSINFOpdf
\else
\fi
\usepackage{graphicx}
\usepackage{amsmath}
\usepackage{array}

\begin{document}

\title{Microcomb-based true-time-delay network for microwave beamforming with arbitrary beam pattern control}

\author{Xiaoxiao~Xue, Yi~Xuan, Chengying~Bao, Shangyuan~Li, Xiaoping~Zheng, Bingkun Zhou, Minghao~Qi, and Andrew~M.~Weiner

\thanks{This work was supported by the National Nature Science Foundation of China (61690191, 61690192, 61420106003), Beijing Natural Science Foundation (4172029), the Air Force Office of Scientific Research (FA9550-15-1-0211), the DARPA PULSE program (W31P40-13-1-0018), and the National Science Foundation (ECCS-1509578).}

\thanks{X.~Xue, S.~Li, X.~Zheng, and B. Zhou are with the Department of Electronic Engineering, Tsinghua University, Beijing 100084, China (e-mail: xuexx@tsinghua.edu.cn; syli@tsinghua.edu.cn; xpzheng@tsinghua.edu.cn; zbk-dee@tsinghua.edu.cn).}
\thanks{Y.~Xuan, C.~Bao, M.~Qi, and A.~M.~Weiner are with the School of Electrical and Computer Engineering, Purdue University, 465 Northwestern Avenue, West Lafayette, Indiana 47907-2035, USA. Y.~Xuan, M.~Qi, and A.~M.~Weiner are also with Birck Nanotechnology Center, Purdue University, 1205 West State Street, West Lafayette, Indiana 47907, USA (email: yxuan@purdue.edu; bao33@purdue.edu; mqi@purdue.edu; amw@purdue.edu).}
}


\maketitle

\begin{abstract}
Microwave phased array antennas (PAAs) are very attractive to defense applications and high-speed wireless communications for their abilities of fast beam scanning and complex beam pattern control. However, traditional PAAs based on phase shifters suffer from the beam-squint problem and have limited bandwidths. True-time-delay (TTD) beamforming based on low-loss photonic delay lines can solve this problem. But it is still quite challenging to build large-scale photonic TTD beamformers due to their high hardware complexity. In this paper, we demonstrate a photonic TTD beamforming network based on a miniature microresonator frequency comb (microcomb) source and dispersive time delay. A method incorporating optical phase modulation and programmable spectral shaping is proposed for positive and negative apodization weighting to achieve arbitrary microwave beam pattern control. The experimentally demonstrated TTD beamforming network can support a PAA with 21 elements. The microwave frequency range is $\mathbf{8\sim20\ {GHz}}$; and the beam scanning range is $\mathbf{\pm 60.2^\circ}$. Detailed measurements of the microwave amplitudes and phases are performed. The beamforming performances of Gaussian, rectangular beams and beam notch steering are evaluated through simulations by assuming a uniform radiating antenna array. The scheme can potentially support larger PAAs with hundreds of elements by increasing the number of comb lines with broadband microcomb generation.
\end{abstract}


\IEEEpeerreviewmaketitle

\section{Introduction}

\IEEEPARstart{M}{icrowave} phased array antennas (PAAs) utilize the interference of electromagnetic waves from multiple sub-antennas to control the beam patterns in free space \cite{ref1}. Compared to mechanically steered antennas, PAAs have many outstanding advantages such as high beam scanning speed, complex beam pattern control, and multibeamforming ability. PAAs are traditionally used in defense applications, and recently attract increasing interest for future 5G wireless communications \cite{ref2}. Traditional analog PAAs rely on microwave phase shifters to control the phase of each sub-antenna. The bandwidth is limited by the beam-squint problem, namely the beam directions of different microwave frequencies diverge at increased steering angles. True-time-delay (TTD) technique (introducing linear phase shift with the frequency) can solve this problem \cite{ref3}. Photonic delay lines emerged as promising candidates for TTD beamforming in terms of propagation loss and bandwidth. A lot of researches can be found in this area \cite{ref4}--\cite{ref47}. Nevertheless, building photonic TTD beamforming networks for practical applications is not a straightforward task. For large arrays which contain a large number of elements, the hardware volume and complexity of photonic TTD systems (including multiple laser sources and tunable TTD units) becomes unfavorable. Several approaches were reported to compress the hardware. One method is based on wavelength-division multiplexing (WDM) and dispersive delay technique \cite{ref24}--\cite{ref31}. The volume of tunable TTD units is largely compressed in such schemes since only one dispersive delay line is required. But the price is increased complexity of the WDM sources which are usually based on laser arrays. Photonic integration techniques are promising to significantly reduce the hardware volume \cite{ref32}--\cite{ref42}. But the bandwidth, delay range, and loss of the state-of-the-art integrated photonic TTD are not yet comparable to those built with discrete components. There are also some TTD methods based on free-space optics \cite{ref43}--\cite{ref47}. Although free-space TTD systems are generally more bulky than fiber based systems, recently reported compact modules show encouraging progresses in reducing the hardware volume \cite{ref45}--\cite{ref47}.

\begin{figure*}[!t]
\centering
\includegraphics[width=6.3in]{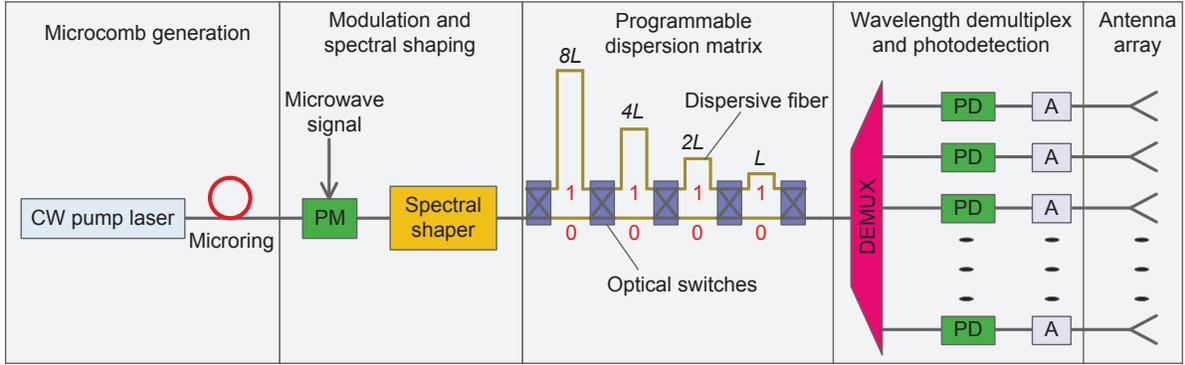}
\caption{Microcomb based microwave TTD beamformer. CW: continuous-wave; PM: optical phase modulator; DEMUX: wavelength demultiplexer; PD: photodetector; A: microwave amplifier.}
\label{fig-1}
\end{figure*}

 Optical frequency combs which contain a series of equally spaced frequency lines are promising candidates for TTD beamforming due to their reduced complexity compared to laser arrays. Previously reported comb sources for such purpose are generated with cascaded or resonant strongly driven electro-optical modulators (i.e., EO combs) \cite{ref_r1}--\cite{ref_r2}. However, the line spacing of EO combs is usually in the range of ten to twenty GHz, corresponding to a limited Nyquist zone (determined by half the comb spacing) typically of a few GHz. Recently, microresonator based optical frequency comb (microcomb) generation has been considered a revolutionary technique and is now investigated intensely for portable applications \cite{ref48}--\cite{ref51}. Microcombs can potentially be fully integrated with CMOS-compatible technology, resulting in an ultra-compact volume \cite{ref52}--\cite{ref53}. The comb line spacing can be up to tens to hundreds of GHz, corresponding to a large Nyquist zone. Moreover, the large bandwidth of microcombs contains a large number of lines which in principle can support large antenna arrays.

 In this paper, we present a detailed investigation of a novel photonic microwave TTD beamforming network based on a miniature microcomb source and a programmable dispersion matrix \cite{ref_r3}. The proposed scheme is capable of arbitrary microwave beam pattern control by employing optical phase modulation and programmable spectral shaping. Limited by the available components in our laboratory, a real antenna array was not constructed; however, detailed measurements of the microwave amplitudes and phases were performed that allow us to simulate the beamforming performance that could be obtained with a real array. The rest of the paper is organized as follows. Section II introduces the scheme of the microcomb-based microwave TTD beamformer, as well as the details of microcomb generation and characterization of the binary dispersion matrix. The problem of higher-order dispersion induced time delay errors and their compensation are discussed. Section III shows the principle of positive and negative weighting for arbitrary beam pattern control, and the simulated beamforming performance based on measurements of the microwave amplitudes and phases. Section IV summarizes the paper. Future work to reduce the whole system volume and improve the performance in a scenario with a real antenna array is also discussed.

\section{Scheme of the microcomb-based microwave TTD beamformer}

The microcomb based microwave TTD beamformer is illustrated in Fig. \ref{fig-1}. A broadband optical frequency comb is generated by pumping a nonlinear microresonator with a continuous-wave laser. The comb is modulated by a microwave signal, shaped by a spectral shaper, and passes through a programmable dispersion matrix. The dispersion matrix is based on a binary delay line constructed of optical switches and dispersive fibers (four bits are illustrated in Fig. \ref{fig-1}) \cite{ref25}. Different comb lines with microwave modulation have different time delays due to the dispersion. The comb lines are then demultiplexed and photodetected. The generated microwave signals are amplified and sent to an antenna array. The beam direction can be controlled by switching the dispersion matrix. And the beam pattern can be controlled by shaping the comb spectrum with the spectral shaper. The operating principle and experimental demonstration of each function is explained in detail in the following sections.

\subsection{On-chip microcomb generation}

\begin{figure}[!t]
\centering
\includegraphics[width=3in]{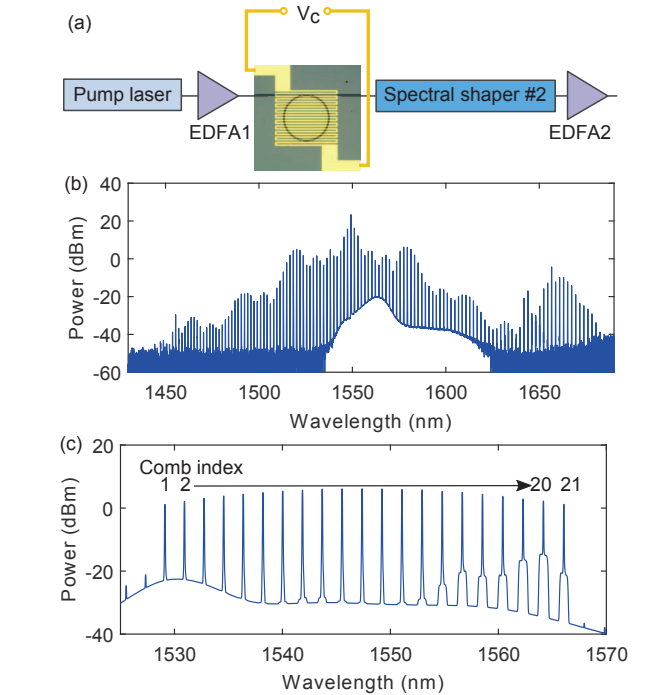}
\caption{(a) Experimental setup of microcomb generation. The nonlinear microresonator is a SiN microring. A spectral shaper (\#2) and an erbium-doped fiber amplifier (EDFA2) are inserted after the microring to smooth the comb spectrum and increase the power. (b) Comb spectrum after the microring. (c) Comb spectrum shaped to a Gaussian window after EDFA2. The resolution of the optical spectrum analyzer is 2.5 GHz (0.02 nm).}
\label{fig-2}
\end{figure}

Figure \ref{fig-2}(a) shows our experimental setup of microcomb generation for TTD beamforming. The nonlinear microresonator is a silicon nitride (SiN) microring fabricated with a CMOS-compatible process. The same device was also used in our previous researches \cite{ref55}--\cite{ref57}. The microring has a loaded quality factor of $\sim8\times 10^5$ , a radius of $100\ \mathrm{\mu m}$, and a free spectral range of $231\ \mathrm{GHz}$. The waveguide constructing the microring has a normal dispersion of $-150\ \mathrm{ps/nm/km}$. A microheater is fabricated on top of the microring. A broadband low-noise comb is generated by fixing the pump laser wavelength and thermally tuning the microring, as described in our previous paper \cite{ref56}. The thermal tuning scheme also offers us the ability of shifting the comb spectrum by tuning the pump laser and the microring in tandem after the comb is generated. This is essential in practical systems to achieve a careful alignment between the comb lines and the wavelength demultiplexing devices.

Figure \ref{fig-2}(b) shows the full-range comb spectrum after the microring. The on-chip pump power is around 28 dBm. The conversion efficiency is more than 30\%, corresponding to an on-chip comb power of 23 dBm excluding the pump. Note that the comb spectrum is not smooth with a fluctuation of $\sim28\ \mathrm{dB}$ in the lightwave C band. In principle, the comb spectral fluctuation can be compensated by programming the spectral shaper after the modulator. However, the spectral shaper we use (Finisar WaveShaper 1000A) has a limited attenuation control range of 35 dB. In that case, it will be impractical to use the same spectral shaper to shape the comb spectrum to arbitrary functions for flexible beam pattern control. Furthermore, the equalization process will introduce a large insertion loss. Therefore, an additional spectral shaper (\#2) is inserted in our experiments to smooth the comb spectrum before modulation; and an erbium-doped fiber amplifier (EDFA) is used to increase the shaped comb power to a total level of 17 dBm. Twenty-one comb lines in the C band are selected and shaped to a truncated Gaussian window. The spectrum is shown in Fig. \ref{fig-2}(c). The normalized power of each comb line is given by
\begin{equation}
p_n = 10^{-(n-11)^2/200}
\end{equation}
where $n=1,2,...,20,21$ the comb index.

\subsection{Characterization of the dispersion matrix and higher-order dispersion compensation}

\begin{figure}[!t]
\centering
\includegraphics[width=3in]{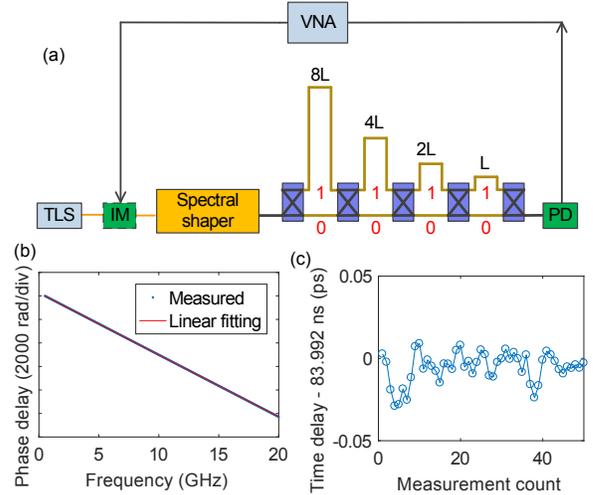}
\caption{(a) Setup for characterizing the dispersion matrix. TLS: tunable laser source; IM: optical intensity modulator; PD: photodetector; VNA: electrical vector network analyzer. (b) Example of measured microwave phase delay versus frequency. (c) Retrieved time delay via linear curve fitting. Fifty measurements are performed with an interval of 1~second.}
\label{fig-3}
\end{figure}

We built the binary dispersion matrix with G.652 single-mode fibers (SMFs) which have a dispersion coefficient between $16\sim17\ \mathrm{ps/nm/km}$ in the C band. The $2\times2$ optical switches are commercial products based on the MEMS technique. The switching time is $<1\ \mathrm{ms}$ and the insertion loss is $<1\ \mathrm{dB}$. The amount of dispersion is directly related to the fiber length. It is therefore critical to measure and cut the fibers accurately to minimize the delay errors. The method we used to determine the fiber length is by measuring the fiber time delay and multiplying by the light speed in fiber. The setup for measuring the time delay is shown in Fig. \ref{fig-3}(a). A microwave signal is generated with a vector network analyzer (VNA) and modulates an optical carrier from a laser source. The laser wavelength is fixed at 1550 nm. The microwave signal is recovered via a photodetector after fiber delay and analyzed by the VNA. The microwave frequency is swept. The time delay is obtained by linearly fitting the microwave phase delay ($\phi$) with the angular frequency ($\omega$), i.e. $\tau=-\mathrm{d}\phi/\mathrm{d}\omega$ (see Fig. \ref{fig-3}(b)). The full frequency sweep range is $0.5\sim20\ \mathrm{GHz}$ with 32001 steps. Smaller fitting errors can be achieved with larger sweep ranges given the same phase measurement errors. Note that the phase change between two consequent frequencies may exceed $2\pi$ when the time delay is large. The phase slope with frequency cannot be resolved correctly in this case. In our experiments, the time delay is first estimated by choosing a smaller sweep range ($10\sim10.5\ \mathrm{GHz}$) and then accurately measured by performing a full-range sweep. We find that this method can achieve a standard deviation of $\sim10\ \mathrm{fs}$ (see Fig. \ref{fig-3}(c)). To measure the light speed in fiber, we picked a short length of fiber ($\sim1\ \mathrm{m}$) which can be measured with a ruler in an accuracy of 1 mm and measured its time delay. The light speed in fiber is then calculated to be $(2.039\pm0.002\times10^{8})\ \mathrm{m/s}$.

\begin{figure}[!t]
\centering
\includegraphics[width=3in]{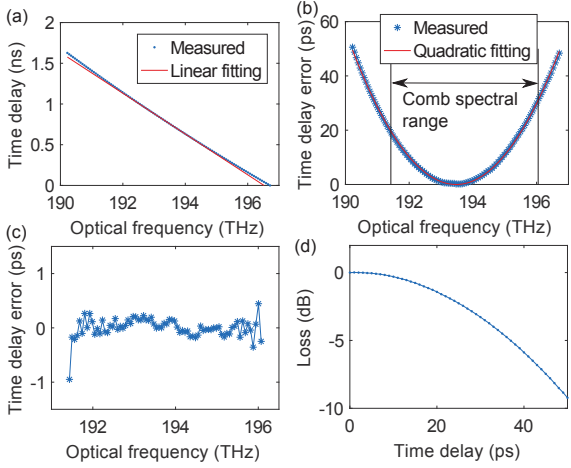}
\caption{(a) Time delay versus the optical frequency when the dispersion matrix is in the ``1111'' state. (b) Time delay errors caused by the third-order dispersion (TOD) compared to an ideal linear function. (c) Time delay errors after TOD compensation with the spectral shaper. (d) Extra loss of the spectral shaper versus its time delay when it is used for delay control.}
\label{fig-4}
\end{figure}

The targeted fiber lengths in the 4-bit dispersion matrix are 1008~m, 504~m, 252~m, and 126~m. The lengths (dispersion) are chosen to support a PAA with a center frequency of 15~GHz and a beam scanning range of $\pm60^\circ$. The spacing between sub-antennas is 1 cm, equal to half of the 2~cm microwave wavelength. The measured fiber lengths are 1007.992~m, 503.995~m, 251.999~m, and 125.998~m. The estimated fiber cutting errors are below 1 cm. However, even with length errors of 1 m, the maximum relative delay error between two adjacent comb lines would be only $\sim30\ \mathrm{fs}$.

We characterized the dispersion matrix by measuring the time delay versus the optical frequency with a tunable laser source. Figure \ref{fig-4}(a) shows the results when the dispersion matrix is in the ``1111'' state. Due to the higher-order dispersion terms of the fiber (mainly the third-order dispersion (TOD)), the relation between the time delay and the optical frequency is not completely linear. By fitting the curve with a second-order polynomial
\begin{equation}
\tau(\omega) = \beta_2 L(\omega-\omega_0)+\beta_3 L/2\cdot(\omega-\omega_0)^2,
\end{equation}
we can retrieve the second- and third-order dispersion coefficients as $\beta_2=-21.08\ \mathrm{ps^2/km}$ and $\beta_3=0.1254\ \mathrm{ps^3/km}$, respectively. Here $L$ is the fiber length and $\omega_0$ is the reference angular frequency which is $2\pi\times194.477\ \mathrm{THz}$. Figure \ref{fig-4}(b) shows the TOD induced delay errors compared to an ideal linear function. Also shown is the microcomb spectral range. The maximum delay error in the comb spectral range is almost 30~ps. For a microwave signal of 15~GHz, it corresponds to a phase error of $162^\circ$. Such a large error will cause severe beam distortions and should be compensated. In our scheme, the TOD induced delay errors can simply be compensated by programming the spectral shaper to apply a phase function that is opposite to the fiber TOD term. The phase function of the spectral shaper is given by
\begin{equation}
\phi(\omega) = \beta_3 L/6\cdot(\omega-\omega_0)^3.
\end{equation}
Figure \ref{fig-4}(c) shows the residual delay errors after TOD compensation. In the comb spectral range, the delay errors are less than 1~ps with a standard deviation of 0.18~ps. It should be noted that the spectral shaper's ability to provide time delay control is limited by its optical resolution (10~GHz in our experiments). An extra loss will be introduced by the spectral shaper when the delay error is large \cite{ref58}. Figure \ref{fig-4}(d) shows the measured extra loss of the spectral shaper versus its time delay. In our demonstration, the maximum delay error caused by higher-order dispersion is around 30 ps (see Fig. \ref{fig-4}(b)). The extra loss of the spectral shaper when compensating this error is about 3~dB. Note that the delay error is a parabolic function in the comb spectral range, corresponding to lower loss in the comb spectral center and higher loss close to the spectral edge. This extra loss function can be taken into account to achieve the overall apodization discussed in the following section. When the maximum delay error becomes considerable for large-scale antenna arrays, additional TOD compensation methods will be required. One solution is combining the dispersive fiber (i.e., SMF in our demonstration) with another type of fiber which has the opposite sign of TOD \cite{ref_r4} to yield overall close-to-zero TOD. Another method is using fiber Bragg gratings which have more compact volume and can be flexibly designed to control both the second- and the higher-order dispersion terms \cite{ref_r5}--\cite{ref_r7}.

\begin{figure}[!t]
\centering
\includegraphics[width=2.5in]{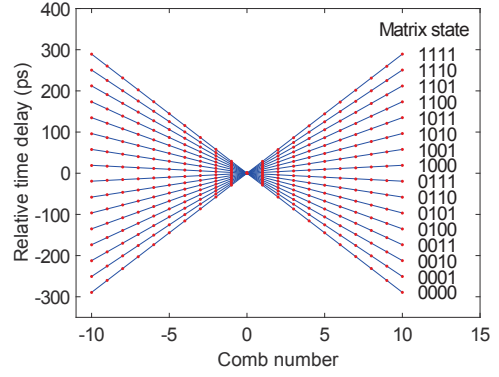}
\caption{Measured time delays of the comb lines shown in Fig. 2(c). The dispersion matrix goes through all the states from ``0000'' to ``1111''. The spectral shaper is used to select each comb line for time delay measurement and compensate the fiber TOD.}
\label{fig-5}
\end{figure}

We then replaced the tunable laser with the microcomb source shown in Fig. \ref{fig-2}(c) and measured the time delay of each comb line. The results are shown in Fig. 5. The dispersion matrix goes through all the states from ``0000'' to ``1111''. The spectral shaper is programmed to a bandpass filter to select each comb line for time delay measurement. For each matrix state, the phase of the spectral shaper is reprogrammed to compensate the fiber TOD. Nice linear delay versus the comb frequency is achieved with a root-mean-square error of $\sim1.1\ \mathrm{ps}$. The larger delay error compared to the estimated value ($<1\ \mathrm{ps}$) is likely due to the fiber non-uniformity and the cross talk from other comb lines in measurement. Note that a fixed delay offset has been removed in Fig. \ref{fig-5}. The relative delay of each comb line in each matrix state is given by
\begin{equation}
\tau_{ijkl}'(n) = \tau_{ijkl}(n) - \left[\tau_{0111}(n)+\tau_{1000}(n)\right]/2 \label{equ-4}
\end{equation}
where $i$, $j$, $k$, $l$ take the value of ``0'' or ``1''. The delay offset adjustment is required to achieve a symmetric beam scanning range with respect to the normal line of a real antenna array. It can be implemented by inserting a length of fiber that has opposite dispersion before the wavelength demultiplexer, such as dispersion-compensating fiber (DCF) commonly used in fiber communications. For a typical DCF with a dispersion coefficient of $-124\ \mathrm{ps/nm/km}$, the required DCF length is estimated to be 126 m. Another method is by adjusting the fiber lengths after the wavelength demultiplexer. The differential fiber length between adjacent channels is about 6~mm. Accurate fiber cutting is thus required to achieve low delay errors. Precise length adjustment can be performed with mechanically tunable fiber-pigtailed delay lines which can have a tuning range of several centimeters \cite{ref59}.

\section{Beam pattern control via spectral shaping}

The far-field amplitude distribution of a PAA is given by
\begin{equation}
E(\theta)\propto \sum_{n=1}^{N} a_n e ^{-\mathrm{j}\left[n\omega_{\mathrm{RF}}d \sin (\theta)/c - \phi_n\right]}
\end{equation}
where $\theta$ is the direction angle relative to the normal line of the PAA; $a_n$ is the tap coefficient of the $n$-th sub-antenna; $\omega_{\mathrm{RF}}$ is the microwave angular frequency; $d$ is the spacing between sub-antennas; $c$ is the light speed in free space; $\phi_n=n\Delta\phi$ is the phase of the $n$-th sub-antenna. The beam direction, i.e. the angle with the maximum microwave amplitude, is given by
\begin{equation}
\theta_{\mathrm{max}} = \arcsin \left( \frac{c\Delta\phi}{\omega_{\mathrm{RF}}d} \right) \label{equ-6}
\end{equation}
For a TTD based PAA as proposed here, the relative phase delay between sub-antennas is given by
\begin{equation}
\Delta\phi = -\omega_{\mathrm{RF}}\Delta\tau \label{equ-7}
\end{equation}
By substituting Eq. (\ref{equ-7}) into Eq. (\ref{equ-6}), we can see that the beam direction can be steered independently of the microwave frequency by controlling the relative time delay between sub-antennas ($\Delta\tau$).The beam shape is the Fourier transform of the tap coefficients. Ideally, both positive and negative taps are required for flexible beam pattern control (e.g. to achieve a rectangular beam). In the following, we demonstrate that arbitrary beam pattern control can be achieved by shaping the comb spectrum and the modulation sidebands with the spectral shaper.

\begin{figure}[!t]
\centering
\includegraphics[width=3.3in]{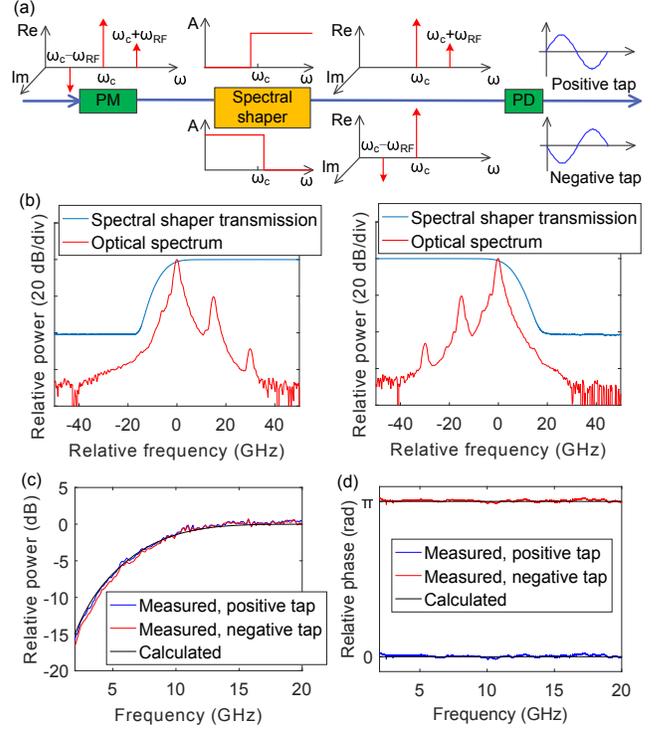}
\caption{(a) Principle of positive and negative tap generation with the spectral shaper. PM: optical phase modulator; PD: photodetector. (b) Optical spectra after the spectral shaper for positive tap (left) and negative tap (right). The transmission of the spectral shaper is also shown. (c) Amplitude and (d) phase responses of the positive and negative taps. The dispersion matrix is in the ``0000'' state. A linear phase related to a constant time delay is removed in the phase response. The measured spectral shaper transmission is used in calculating the theoretical results.}
\label{fig-6}
\end{figure}

\begin{figure}[!t]
\centering
\includegraphics[width=3.2in]{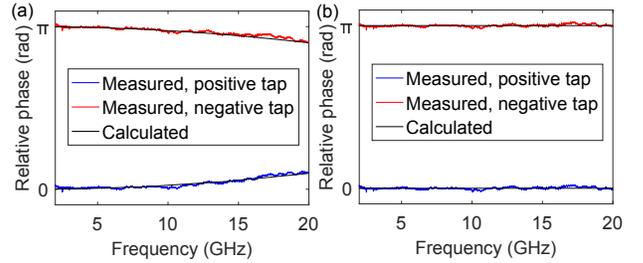}
\caption{Microwave phase responses when the dispersion matrix is in the ``1111'' state. (a) Without local dispersion compensation. (b) With local dispersion compensation.}
\label{fig-7}
\end{figure}

The principle of generating positive and negative taps is shown in Fig. \ref{fig-6}(a). As has been illustrated in Fig. \ref{fig-1}, an optical phase modulator is used to modulate the frequency comb. The optical field after modulation for each comb line is given by
\begin{equation}
e(t) = A e^{\mathrm{j}\left[\omega_{\mathrm{c}}t+m\sin (\omega_{\mathrm{RF}}t)\right]}
\end{equation}
where $A$ is the field amplitude normalized that $A^2$ represents the power; $\omega_{\mathrm{c}}$ is the comb angular frequency; $\omega_{\mathrm{RF}}$ is the microwave frequency; and $m$ is the modulation index given by $m=\pi V/V_\pi$ where $V$ is the microwave amplitude and $V_\pi$ is the modulator half-wave voltage. Under small signal approximation, the optical field contains three terms:
\begin{align}
\mathrm{carrier:\ } &e_{\mathrm{c}} = A e^{\mathrm{j}\omega_{\mathrm{c}}t}, \\
\mathrm{upper\ sideband:\ } &e_{\mathrm{usb}} = \frac{mA}{2}e^{\mathrm{j}(\omega_{\mathrm{c}}+\omega_{\mathrm{RF}})t}, \\
\mathrm{lower\ sideband:\ } &e_{\mathrm{lsb}} = -\frac{mA}{2}e^{\mathrm{j}(\omega_{\mathrm{c}}-\omega_{\mathrm{RF}})t},
\end{align}
One sideband is then suppressed by programming the amplitude of the spectral shaper. By selectively suppressing the lower or upper sideband, positive and negative taps can be generated. When the lower sideband is suppressed, the microwave signal generated in the photodetector is given by
\begin{equation}
V_{\mathrm{RF}}^{+} = \Re \cdot R_{\mathrm{L}} \cdot e_{\mathrm{usb}} \cdot e_{\mathrm{c}}^* = \frac{m\Re R_{\mathrm{L}} A^2}{2} e^{\mathrm{j}\omega_{\mathrm{RF}}t} \label{equ-11}
\end{equation}
where $\Re$ is the responsivity of the photodetector and $R_{\mathrm{L}}$ is the load resistance. When the upper sideband is suppressed, the microwave signal generated in the photodetector is given by
\begin{equation}
V_{\mathrm{RF}}^{-} = \Re \cdot R_{\mathrm{L}} \cdot e_{\mathrm{c}} \cdot e_{\mathrm{lsb}}^* = -\frac{m\Re R_{\mathrm{L}} A^2}{2} e^{\mathrm{j}\omega_{\mathrm{RF}}t} \label{equ-12}
\end{equation}
Figure \ref{fig-6}(b) shows the measured optical spectra of one comb line after the spectral shaper. Figure \ref{fig-6}(c) shows the microwave amplitude. When the microwave frequency is low, the sideband cannot be suppressed efficiently due to the limited resolution of the spectral shaper \cite{ref58}. Therefore, the microwave amplitude shows a high-pass response. The 3-dB lower cut-off frequency is 8~GHz. Figure \ref{fig-6}(d) shows the microwave phase. A clear $\pi$-phase difference can be observed between the positive and negative taps.

Note that the dispersion matrix is in the ``0000'' state for the measurements performed in Fig. \ref{fig-6}. The optical dispersion is negligible and not considered in deriving Eqs. (\ref{equ-11}) and (\ref{equ-12}). When the dispersion is large, e.g. with the dispersion matrix in the ``1111'' state, the optical dispersion is transferred to the electrical domain and causes microwave phase errors. The microwave signals related to positive and negative taps are given by
\begin{align}
V_{\mathrm{RF}}^{+} &=  \frac{m\Re R_{\mathrm{L}} A^2}{2} e^{\mathrm{j}\omega_{\mathrm{RF}}t} e^{-\mathrm{j}\beta_2L\omega_{\mathrm{RF}}^2/2}, \\
V_{\mathrm{RF}}^{-} &=  -\frac{m\Re R_{\mathrm{L}} A^2}{2} e^{\mathrm{j}\omega_{\mathrm{RF}}t} e^{\mathrm{j}\beta_2L\omega_{\mathrm{RF}}^2/2}.
\end{align}
Here we omit the TOD since it is much smaller than the second-order dispersion. Figure \ref{fig-7} shows the measured microwave phase when the dispersion matrix is in ``1111'' state. A clear deviation from ``0'' and ``$\pi$'' can be observed at high frequencies. The electrical phase distortion can be compensated by local dispersion compensation with the spectral shaper. The phase of the spectral shaper is programmed as
\begin{equation}
\phi(\omega) = \beta_2L/2\cdot(\omega-\omega_c)^2,
\end{equation}
where $\omega_{\mathrm{c}}$ is the comb angular frequency. Figure \ref{fig-7}(b) shows the measured microwave phase after local dispersion compensation. Note that by combining the global TOD compensation introduced in last section, the overall phase of the spectral shaper around the comb line $\omega_{\mathrm{c}}$ should be given by
\begin{equation}
\phi(\omega) = \beta_2L/2\cdot(\omega-\omega_{\mathrm{c}})^2 + \beta_3L/6\cdot(\omega-\omega_0)^3.
\end{equation}

\begin{figure*}[!t]
\centering
\includegraphics[width = 6.7in]{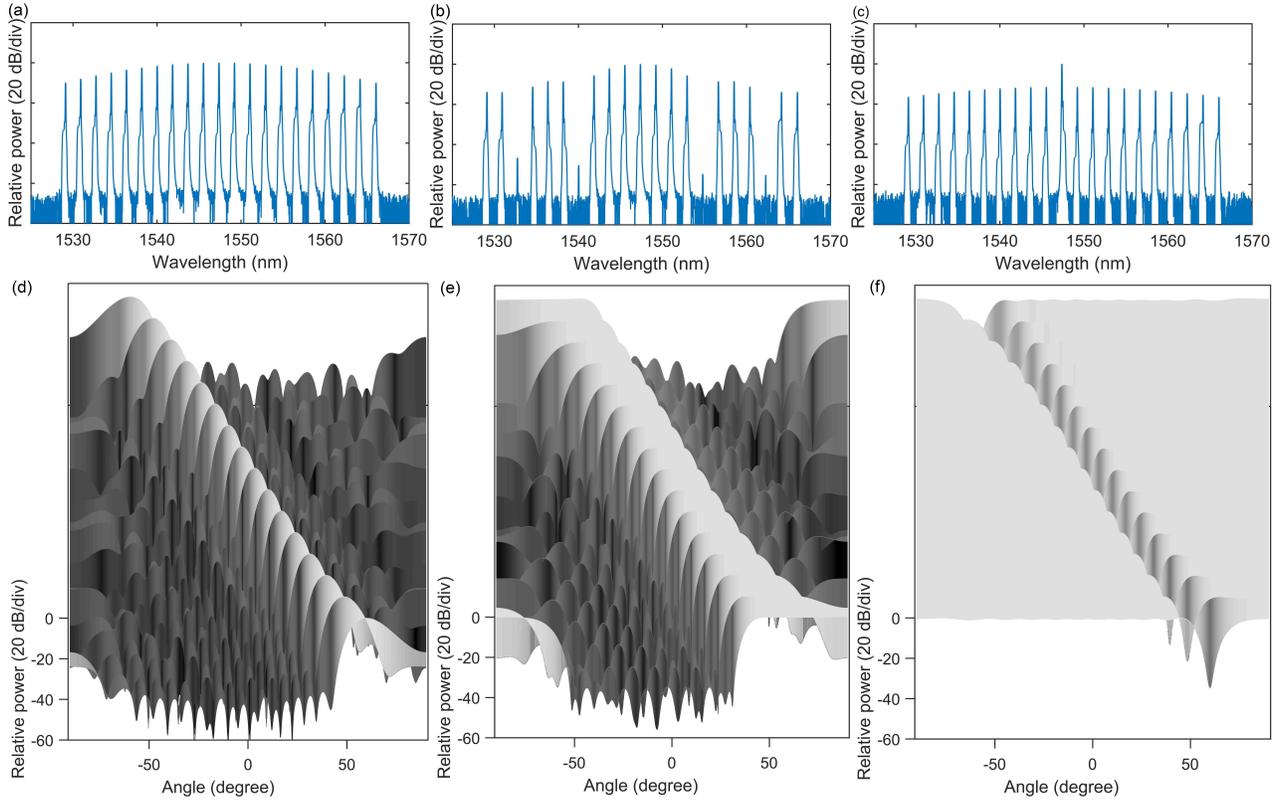}
\caption{Beam pattern control with optical spectral shaping. (a), (b), (c) Optical spectra after the spectral shaper for Gaussian beam, rectangular beam, and dark beam, respectively. (d), (e), (f) Calculated beam patterns based on measured microwave amplitude and phase, for Gaussian beam, rectangular beam, and dark beam, respectively. The microwave frequency is 15 GHz.}
\label{fig-8}
\end{figure*}

Limited by the available components in our laboratory, we did not build the photodetector array, microwave amplifier array, and the antenna array illustrated in Fig. \ref{fig-1}. To evaluate the beamforming ability, we measured the microwave amplitude and phase corresponding to each sub-antenna (comb line), and calculated the beam patterns by assuming identical photodetector and amplifier characteristics. The dispersion matrix goes through all the states from ``0000'' to ``1111''. The microwave amplitude and phase corresponding to each comb line in each matrix state are measured ($21\times16$ measurements in total for a full scan test; frequency sweep with the VNA is performed for each measurement). By programming the spectral shaper after the optical modulator, the tap coefficients are first shaped to a Gaussian function given by
\begin{equation}
a_n = 10^{-(n-11)^2/100}.\label{equ-18}
\end{equation}
The optical spectrum after the spectral shaper is shown in Fig. \ref{fig-8}(a). With an iterative adjusting strategy, shaping errors less than 0.2~dB can be achieved. The calculated beam patterns are shown in Fig. \ref{fig-8}(d). A fixed delay offset between the sub-antennas has been assumed in the calculation (see Eq. (\ref{equ-4}) and related discussions). The microwave frequency is 15~GHz. By switching the dispersion matrix, the beam can scan from $-60.2^\circ$ to $60.2^\circ$ with an average step of $8^\circ$. The 3-dB beam width is $7^\circ$ at the angles of $\pm3.3^\circ$ and $14^\circ$ at the angles of $\pm60.2^\circ$. The out-of-beam suppression is larger than 30~dB.

To demonstrate the ability of beamforming with negative taps, the tap coefficients are then shaped to a sinc function given by
\begin{equation}
a_n = 10^{-(n-11)^2/200}\cdot \frac{\sin\left[(n-11)\pi/4\right]}{(n-11)\pi/4}.
\end{equation}
The optical spectrum is shown in Fig. \ref{fig-8}(b). The microwave amplitudes and phases are measured again for a full-range beam scan. The calculated beam patterns are rectangular as shown in Fig. \ref{fig-8}(e). The 3-dB beam width is about $25^\circ$ at the angles of $\pm3.3^\circ$ and $50^\circ$ at the angles of $\pm60.2^\circ$.

The microwave notch direction can also be controlled with the TTD beamforming scheme. Here we demonstrate a ``dark'' beam, i.e. an omnidirectional beam with a notch at a specified angle. Figure \ref{fig-8}(c) shows the optical spectrum. The tap coefficients are given by
\begin{equation}
a_n =
\begin{cases}
-14, & n=11 \\
10^{-(n-11)^2/200}, & n=1,...,10,12,...,21
\end{cases}
\end{equation}
The 11-th tap has an opposite sign compared to the other taps. Figure \ref{fig-8}(f) shows the calculated beam patterns. A notch with a suppression ratio of $\sim25\ \mathrm{dB}$ can be clearly observed. The null direction is where the amplitude of the omnidirectional beam generated by the negative tap is cancelled by the positive beam generated by the other taps. Note that we did not optimize the notch suppression ratio in this demonstration. Higher suppression ratios may be achieved by optimizing the relative power level of the negative tap (i.e. the 11-th comb line). The ability of null steering can be used for minimizing the interferences between multiple users and targets \cite{ref60}.

\begin{figure}[!t]
{\centering
\includegraphics[width = 3.2in]{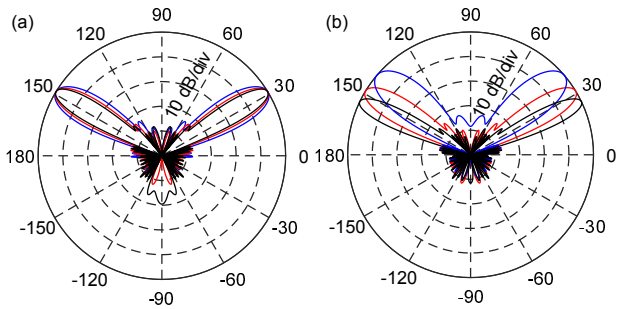}
\caption{Calculated beam patterns of 3 microwave frequencies with (a) our TTD based beamformer and (b) a phase shifter based beamformer. Blue: 12~GHz. Red: 15~GHz. Black: 18~GHz.}
\label{fig-9}}
\end{figure}

Finally, to prove that our TTD beamforming scheme is free of the beam-squint problem, we calculated the beam patterns for three frequencies, i.e. 12~GHz, 15~GHz, and 18~GHz. The beam shape is Gaussian and the beam direction is $31.4^\circ$. The corresponding tap coefficients are given by Eq. (\ref{equ-18}); and the dispersion matrix state is ``0011''. For comparison, we also calculated the beam patterns for a phase shifter controlled array. The phases of all the frequencies in the phased array are assumed the same as that of the 15 GHz. The results are shown in Fig. \ref{fig-9}. For the TTD-based scheme, no obvious beam divergence is observed. By contrast, there is a large beam divergence for the phase shifter based scheme. The beam directions of 12~GHz, 15~GHz, and 18~GHz are $40.4^\circ$, $31.4^\circ$, and $25.6^\circ$, respectively. Therefore, our TTD based PAA can support a much larger microwave bandwidth than a phase shifter based PAA.

\section{Summary and discussion}
In summary, we have demonstrated a miniature microcomb based TTD beamforming network with the ability of arbitrary beam pattern control. The beamforming network can support a PAA with 21 elements. The microwave frequency range is $8\sim20\ \mathrm{GHz}$ (upper frequency limited by the measurement equipment); and the beam scanning range is $\pm60.2^\circ$. By increasing the number of comb lines, the proposed scheme can potentially support large-scale PAAs. For example, with 50~GHz comb line spacing, the comb lines in S, C and L bands will be able to support more than 400~sub-antennas. The use of a microcomb for photonic microwave TTDs was also proposed recently in \cite{ref_r8}. A ``semi-coherent'' type-II comb with 49~GHz spacing was shown. Although high coherence is not essential for photonic microwave TTDs in principle, type-II combs are usually accompanied by high intensity noise which will compromise the radiofrequency performance \cite{ref_r9}--\cite{ref_r11}. There were also large variations in the comb spectrum, which reduces efficiency after apodization. Therefore, efforts are still required to develop coherent broadband 50 GHz microcombs for practical applications.

Our experiments have shown that microcombs are promising candidates for microwave TTD beamforming as compact chip-level integrated multi-wavelength sources. There are still several ways to further reduce the volume of the whole beamforming system.  One attractive solution is integrating the pump laser with the microresonator. Novel methods to increase the laser output power and reduce the pump threshold are critical to achieve this goal. It is also important to increase the power conversion efficiency and reduce the comb spectral variation, so that additional optical amplification and equalization may not be necessary. In our current setup, EDFAs are used to increase the power of the pump and the comb. Amplified spontaneous emission (ASE) noise is introduced. The optical signal-to-noise ratio (SNR) is $11\sim31\ \mathrm{dB}$ for different channels assuming the standard 10~GHz bandwidth (see Fig. \ref{fig-2}(c) with a measurement spectral resolution of 2.5~GHz). The signal-ASE beat noise constitutes the main microwave noise source. For a modulator half-wave voltage of 3~V and a microwave modulation power of 10~dBm, the resulting electrical SNR with an electrical bandwidth of 2.5~GHz is between $10\sim30\ \mathrm{dB}$. High-efficiency microcomb generation without EDFAs is expected to greatly reduce the noise.

Programmable spectral shapers are key elements in our setup to perform apodization and fine time delay tuning. The commercially available spectral shaper based on liquid crystal on silicon technique has reasonably compact volume \cite{ref_r12}. Efforts are underway in the research community to develop chip-scale integrated spectral shapers \cite{ref_r13}--\cite{ref_r17}. Another way to further reduce the system volume is building the dispersion matrix with highly dispersive components such as photonic crystal fibers \cite{ref_r18}, chirped fiber Bragg gratings \cite{ref_r19}, or integrated waveguides \cite{ref_r20}. Such devices may achieve dispersion coefficients tens to hundreds of times that of SMFs, thus require significantly shorter lengths to produce the same amount of dispersion.

Finally, the beamforming performances of our scheme are evaluated through simulations by assuming a uniform PAA in which the characteristics of the components in each channel are identical. In a real antenna array, there are inevitably differences in the characteristics of the photodetectors, microwave amplifiers, and sub-antennas. The non-uniformity would cause distortions of the microwave beam shape and degradation of the mainlobe-to-sidelobe suppression. In our current demonstration, optical spectral shaping including both amplitude and phase control has been used to apodize the amplitudes of the array elements for arbitrary beam pattern control as well as to compensate phase errors connected to the dispersion. It can also be helpful to compensate the channel non-uniformity in a practical PAA. Such a photonic TTD beamformer involving a real antenna array will be investigated in our future work.


\ifCLASSOPTIONcaptionsoff
  \newpage
\fi

%

\begin{IEEEbiographynophoto}{Xiaoxiao Xue}
received the B.S. and Ph.D. degrees in electronic engineering with the highest honors from Tsinghua University, Beijing, China, in 2007 and 2012, respectively. He was a recipient of the 2012 Wang Daheng Prize funded by the Optical Society of China for his Ph.D. dissertation on microwave photonic signal processing. From 2013 to 2015, he worked as a postdoctoral researcher in the Ultrafast Optics and Optical Fiber Communications Laboratory in Purdue University. Since 2016, he joined the Department of Electronic Engineering in Tsinghua University as an assistant professor. He has published more than 60 journal and conference papers. His research interests include optical frequency comb generation, microwave photonic signal processing, radio over fiber, and phased array antennas.
\end{IEEEbiographynophoto}

\begin{IEEEbiographynophoto}{Yi Xuan}
received the B.S. and M.S. degrees in chemistry from the East China University of Science and Technology, Shanghai, China, in 1994 and 1997, respectively, and the Ph.D. degree in inorganic materials from the Tokyo Institute of Technology, Tokyo, Japan, in 2001. Between 2001 and 2005, he was a Postdoctoral Researcher with the National Institute for Materials Science and the National Institute of Advanced Industrial Science and Technology, Tsukuba, Japan. Since August 2005, he has been a Researcher with the School of Electrical and Computer Engineering, Purdue University, West Lafayette, IN, where he is involved in research on nano-CMOS and photonics.
\end{IEEEbiographynophoto}

\begin{IEEEbiographynophoto}{Chengying Bao}
biography unavailable.
\end{IEEEbiographynophoto}

\begin{IEEEbiographynophoto}{Shangyuan Li}
biography unavailable.
\end{IEEEbiographynophoto}

\begin{IEEEbiographynophoto}{Xiaoping Zheng}
biography unavailable.
\end{IEEEbiographynophoto}

\begin{IEEEbiographynophoto}{Bingkun Zhou}
biography unavailable.
\end{IEEEbiographynophoto}

\begin{IEEEbiographynophoto}{Minghao Qi}
received the B.S. degree from the University of Science and Technology of China, Anhui, China, and the M.S. and Ph.D. degrees in electrical engineering from the Massachusetts Institute of Technology, Cambridge, in 2005. In 2005, he joined Purdue University, West Lafayette, IN, where he is currently leading the Photonics and Nano-Fabrication Lab, School of Electrical and Computer Engineering. He is also with the Shanghai Institute of Microsystem and Information Technology, Chinese Academy of Science, Shanghai, China.
\end{IEEEbiographynophoto}

\begin{IEEEbiographynophoto}{Andrew M. Weiner}
graduated from M.I.T. in 1984 with an Sc.D. degree in electrical engineering. Upon graduation he joined Bellcore, first as a Technical Staff Member and later as a Manager of Ultrafast Optics and Optical Signal Processing Research. He moved to Purdue University in 1992 and is currently the Scifres Family Distinguished Professor of Electrical and Computer Engineering. His research focuses on ultrafast optics signal processing and applications to high-speed optical communications and ultrawideband wireless. He is especially well known for his pioneering work on programmable femtosecond pulse shaping using liquid crystal modulator arrays. He is the author of a textbook entitled Ultrafast Optics and has published more than 300 journal articles. He is a Fellow of the OSA, a member of the U.S. National Academy of Engineering and National Academy of Inventors, and National Security Science and Engineering Faculty Fellow. He has won numerous awards for his research, including the Hertz Foundation Doctoral Thesis Prize, the OSA Adolph Lomb Medal, the ASEE Curtis McGraw Research Award, the International Commission on Optics Prize, the IEEE LEOS William Streifer Scientific Achievement Award, the Alexander von Humboldt Foundation Research Award for Senior U.S. Scientists, the OSA R.W. Wood Prize, and the IEEE Photonics Society Quantum Electronics Award. He has served as the Chair or Co-Chair of the Conference on Lasers and Electro-Optics, the International Conference on Ultrafast Phenomena and the National Academy of Engineering¡¯s Frontiers of Engineering symposium, as the Secretary/Treasurer of the IEEE Lasers and Electro-optics Society (LEOS), and as the Vice-President of the International Commission on Optics (ICO). He is currently the Editor-in-chief of Optics Express.
\end{IEEEbiographynophoto}





\end{document}